\documentclass[aps,prl,twocolumn,amsmath,amsart,amssymb,superscriptaddress]{revtex4-1}

\usepackage[latin9]{inputenc}

\usepackage{amsmath}
\usepackage{graphicx}

\usepackage{times}
\usepackage{ulem}

\usepackage{color}

\newcommand{\FSS}{FeSe$_{1-x}$S$_{x}$}

\begin{document}

\title{Nematic charge-density-wave correlations in {\FSS}}

\author{Ruixian Liu$^\#$}
\affiliation{Center for Advanced Quantum Studies and Department of Physics, Beijing Normal University, Beijing, 100875 P. R. China}

\author{Wenliang Zhang$^\#$}
\author{Yuan Wei}
\affiliation{Photon Science Division, Swiss Light Source, Paul Scherrer Institut, CH-5232 Villigen PSI, Switzerland}

\author{Zhen Tao}
\affiliation{Center for Advanced Quantum Studies and Department of Physics, Beijing Normal University, Beijing, 100875 P. R. China}
\affiliation{Photon Science Division, Swiss Light Source, Paul Scherrer Institut, CH-5232 Villigen PSI, Switzerland}

\author{Teguh C. Asmara}
\affiliation{Photon Science Division, Swiss Light Source, Paul Scherrer Institut, CH-5232 Villigen PSI, Switzerland}

\author{Vladimir N. Strocov}

\author{Thorsten Schmitt}
\email{thorsten.schmitt@psi.ch}
\affiliation{Photon Science Division, Swiss Light Source, Paul Scherrer Institut, CH-5232 Villigen PSI, Switzerland}

\author{Xingye Lu}
\email{luxy@bnu.edu.cn}
\affiliation{Center for Advanced Quantum Studies and Department of Physics, Beijing Normal University, Beijing, 100875 P. R. China}

\begin{abstract}
The occurrence of charge-density-wave (CDW) order is a common thread in the phase diagram of cuprate high-transition-temperature ($T_c$) superconductors. In iron-based superconductors (FeSCs), nematic order and fluctuations play a decisive role in driving other emergent orders. CDW order has been observed by scanning tunneling microscopy for various FeSCs such as FeSe thin films, uniaxially strained LiFeAs, and tetragonal FeSe$_{0.81}$S$_{0.19}$. However, it remains elusive if the CDW in these materials is a bulk phenomenon as well as if and how it intertwines with the electronic nematicity. Using energy-resolved resonant X-ray scattering at the Fe-L$_3$ edge, we report the discovery of a local-strain-induced incommensurate isotropic CDW order in FeSe$_{0.82}$S$_{0.18}$. A highly anisotropic CDW response under uniaxial strain unambiguously manifests that the CDW is directly coupled to the nematicity. Transforming part of Fe$^{2+}$ to Fe$^{3+}$ on the surface of {\FSS} reveals that the same isotropic CDW can be induced, enhanced, and stabilized in the whole nematic regime measured ($x=0-0.19$). As Fe$^{3+}$ can create local lattice distortions on the surface, the CDW could arise from the interaction between the local strain around Fe$^{3+}$ and the nematic electron correlations. Our experimental observation of a local-strain-induced CDW gives vital information for understanding the interplay between electron correlations and the electronic nematicity in FeSCs.
\end{abstract}

\maketitle


The intertwining of multiple orders has been suggested to be essential for understanding the microscopic mechanism of unconventional superconductivity in high-$T_c$ superconductors \cite{Fradkin2015, Fernandes2022}. Charge-density wave (CDW) order, defined as itinerant (or valence) electrons/holes organizing into periodical spatial modulating patterns with reduced translational symmetry, has been demonstrated to be a universal feature of various unconventional superconductors such as cuprates, and nickel-based systems \cite{Comin2016, Frano2020, Ghiringhelli2012, Chang2012, Comin2014, Neto2014, Gerber2015, Kim2018, Arpaia2019, Neto2015, Neto2016, Comin2019, Tam2022, Rossi2022, Krieger2022, Lee2019, BNA, Lee2021}. 

In FeSCs, a ubiquitous electronic nematicity (characterized by reduction of rotational symmetry from $C_4$ to $C_2$ driven by an electronic instability) intertwines with superconductivity and the other emergent orders and therefore has attracted substantial research interest in the past decade \cite{Fernandes2014, Chu2010, Yi2011, Chu2012, Kuo2016, Lu2014}.  
Compared with the CDW in cuprates and nickelates confirmed by {\it bulk}-sensitive probes such as resonant elastic x-ray scattering (REXS), various possible CDWs in FeSCs were observed by surface-sensitive scanning tunneling microscopy (STM) \cite{Li2017, Li2021, LFA2018, LD2021, Li2022}, but have rarely been confirmed by bulk probes. For instance, STM measurements found two-fold symmetric ``stripe-like'' smectic order (with a wave vector $q_{\rm smc} \sim 0.2-0.3 \cdot 2\pi/a$ directed along the orthorhombic $a/b$ axis) in FeSe thin films (with a period $\lambda\sim2$ nm) and uniaxial-strained LiFeAs (with $\lambda\sim2.7$ nm) \cite{Li2017, Li2021, LFA2018, LD2021, Li2022}, which were suggested to be manifestations of nematicity at the surface \cite{Lahiri2022}. On the other hand, more recent STM measurements on a tetragonal FeSe$_{0.81}$S$_{0.19}$ sample revealed a stripe CDW exhibiting local $C_2$ symmetry with $q\approx0.12$ {\AA}$^{-1}$ ($\lambda\sim5$ nm) and suggested that rotational symmetry breaking of this CDW is distinct from that of the nematic state in FeSe \cite{Walker2023}.
Therefore, the crucial question remaining is if CDW in FeSCs may also exist in the bulk of these materials and how it intertwines with the electronic nematicity.

\begin{figure*}
\includegraphics[width=16cm]{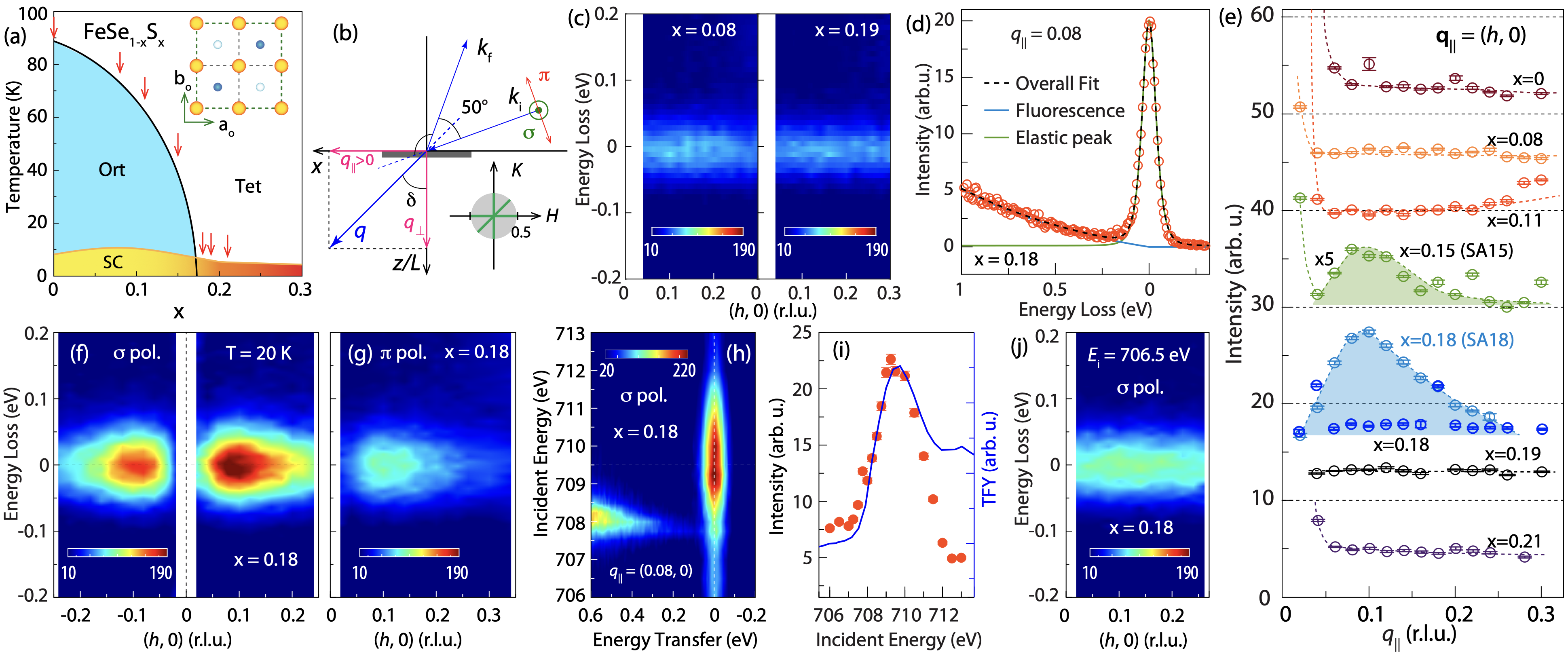}
\caption{\textbf{Phase diagram, scattering geometry, and CDW of {\FSS}}. 
(a), Electronic phase diagram of {\FSS}, where Ort, SC, and Tet denote orthorhombic, superconducting, and tetragonal phases, respectively. The vertical arrows mark the doping levels $x=0, 0.08, 0.11, 0.15, 0.18, 0.19$ and $0.21$ of the investigated samples. 
The inset shows the in-plane lattice structure of {\FSS}. The yellow-filled circles mark Fe ions and the filled and open blue circles denote Se/S ions. The green dashed square denotes the orthorhombic unit cell. (b), Scattering geometry for RIXS measurements. The scattering angle is set to $2\theta_s=130^\circ$. The right, lower inset exhibits the reciprocal space covered by Fe-L$_3$ RIXS (gray filled circle). 
(c) $q_\shortparallel$-dependent elastic scattering along the $H$ axis of freshly cleaved $x=0.08$ and $0.19$ crystals, measured at Fe-L$_3$ edge with $\sigma$ polarization and $T=15$ K. (d), Fit of the elastic peak of RIXS spectrum measured at $\mathbf{q}_\shortparallel = (0.08, 0)$. The black dashed curve is the overall fit and the blue and green solid curves are the fittings of the fluorescence background and the elastic scattering, respectively. (e), $q_\shortparallel$-dependent elastic scattering along $H$ for {\FSS}. The data for $x = 0, 0.08, 0.11, 0.18$ (blue circles), $0.19$ and $0.21$ were collected on freshly-cleaved samples, while the $x=0.15$ and $0.18$ (cyan circles) were measured on aged samples exposed to the ultra-high vacuum in the chamber for $\sim20$ hours. (f), (g), $q_\shortparallel$-dependent elastic scattering of the aged $x=0.18$ sample measured with (f) $\sigma$ and (g) $\pi$ incident light polarizations. (h), Incident-energy dependent RIXS the aged $x=0.18$ sample with $q_\shortparallel = (0.08, 0)$, measured near the Fe-L$_3$ edge with $\sigma$ polarization. 
(i), Comparison between the total-fluorescence yield (TFY) X-ray absorption spectrum (blue curve) and the incident energy dependence of the integrated intensity (red dots) of the elastic scattering of $q_\shortparallel=(0.08, 0)$ as shown in (h). 
(j), Same measurements as (f) performed at $E_i=706.5$ eV. All the measurements were performed at $T = 15-20$ K.
}
\label{fig1}
\end{figure*}

{\FSS} is a unique system for investigating the intertwining between electronic nematicity and other emergent orders among FeSCs because of its simple structure, the absence of static spin order below the structural (nematic) transition, and the pristine nematic quantum critical point (NQCP) (Fig. 1(a))
 \cite{Coldea2021, Hosoi2016, Licciardello2019, Culo2021, Sato2018}.  {\FSS} exhibits a tetragonal-to-orthorhombic structural (nematic) transition at $T_s$, below which the electronic nematic order is defined (Fig. 1(a)). The nematic transition is gradually suppressed upon sulfur doping from $T_s$ = 90 K for FeSe to $T_s$ = 0 at the putative NQCP $x_c\approx0.17$. Concomitantly, the superconducting transition temperature $T_c$ increases from $T_c$ = 8 K to optimal $T_c\approx10$ K at $x\approx0.08$, and decrease to $T_c\approx5$ K across the NQCP \cite{Coldea2021}. 
As various CDWs were observed in FeSe and {\FSS}, it is natural to choose {\FSS} hosting a tunable electronic nematicity (by sulfur doping and uniaxial strain) to explore the presence of a bulk CDW and its interplay with the electronic nematicity.

In this work, we use resonant inelastic x-ray scattering (RIXS) to measure the elastic scattering of unstrained {\FSS} ($x=0-0.21$) and uniaxial-strained FeSe$_{0.82}$S$_{0.18}$ samples covering the regime of nematic order and NQCP (Fig. 1(a)). RIXS measurements on fresh surfaces of in-situ cleaved {\FSS} single crystals at $T=15-20$ K reveal no sign of CDW along $[H, 0]$. However, an isotropic two-dimensional (2D) CDW ($C_{\rm inf}$ in $[H, K]$) with $|q_\shortparallel|\approx0.10$ emerges in an $x=0.18$ sample after being exposed to ultra-high vacuum ($\sim10^{-10}-\sim10^{-9}$ mbar) at $T=15-20$ K for $\sim20$ hours (aging effect). This isotropic CDW in $x=0.18$ is driven to be $C_2$ symmetric under compressive uniaxial strain, indicating that it is directly coupled to the electronic nematicity. Moreover, through inducing a small amount of Fe$^{3+}$ states on the surface at higher temperature ($T\gtrsim80$ K), the isotropic CDW emerges in the whole nematic regime studied ($x=0-0.19$). As defects generated by aging of the surface and Fe$^{3+}$ could induce local strains near the surface, we attributed the emergence and the enhancement of this CDW to a local-strain-induced electron-density modulations in the nematic regime. 
Our results are consistent with the theoretical scenario concerning the origin of the smectic order \cite{Lahiri2022} and provide new insights into understanding the effects of electronic nematicity on the electronic properties in FeSCs.

\begin{figure}[htbp!]
\includegraphics[width=8 cm]{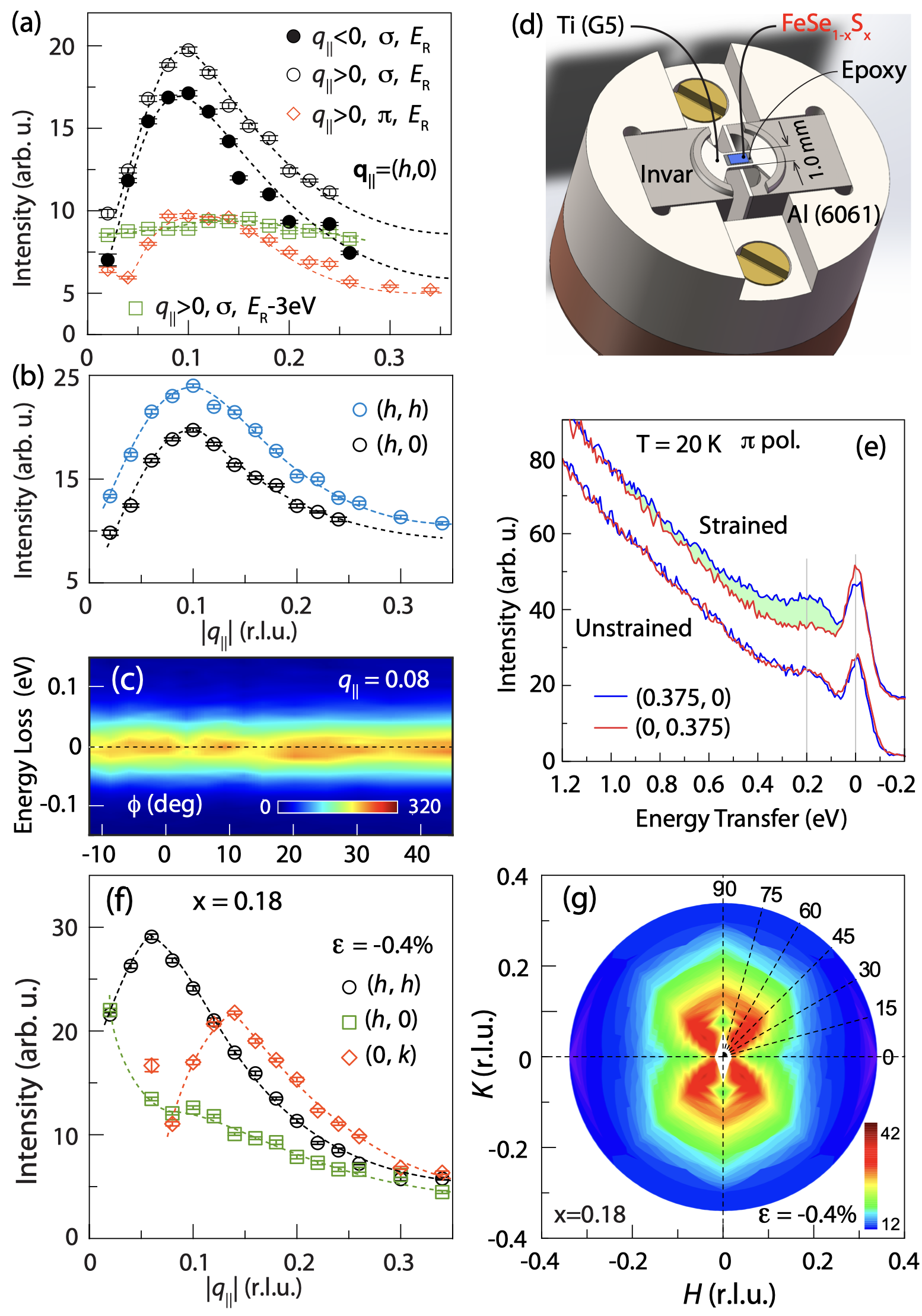}
\caption{\textbf{Uniaxial-strain effects on the CDW in FeSe$_{0.82}$S$_{0.18}$.} (a), CDW peak along the $H$ axis. The data points are obtained from the fitting of the elastic scattering as shown in Fig. 1. All the measurements were performed with $\sigma$ polarization, $E_R$ = 709.5 eV and positive $q_\shortparallel$ unless otherwise stated. The black-filled circles are data measured in $q_\shortparallel<0$ region, the red diamond data with $\pi$ polarization, and the green square data at the pre-edge $E_i = E_R-3 = 706.5$ eV. (b), CDW peaks measured in the SA18 sample along the $H$ and $[H, H]$ directions. (c), Azimuthal dependence of the elastic scattering for $q_\shortparallel=(0.08, 0)$ measured on SA18. (d), Uniaxial-strain device based on differential thermal expansion of aluminum and invar alloy. (e), RIXS spectra with $q_\shortparallel = (0.375, 0)$ and $(0, 0.375)$ measured on strained and unstrained SA18. The green-colored area marks the spin-excitation difference between $q_\shortparallel = (0.375, 0)$ and $(0, 0.375)$. (f), CDW peaks measured in strained SA18 sample along the $H/K$ and $[H, H]$ directions. (g), Azimuthal dependence of the CDW in the reciprocal space of the strained SA18 sample. }
\label{fig2}
\end{figure}

Fig. 1(b) shows the scattering geometry usually used in RIXS studies of FeSCs, and the accessible area of Fe-L$_3$ (2$p_{3/2}-3d$ transition, $h\nu\approx709$ eV) RIXS in the reciprocal space of {\FSS} \cite{Lu2022}.  We first measure the momentum-dependent elastic scattering of {\FSS} with $x=0, 0.08, 0.11, 0.18, 0.19,$ and $0.21$. On these freshly-cleaved {\FSS} single crystals, momentum-dependent colormaps of elastic scattering measured with $\sigma$ polarized incident photons at the Fe-L$_3$ resonance (on-resonance $q_\shortparallel$ scan) are featureless, implying the absence of a 2D CDW along the $H$ axis (Fig. 1(c)). To quantify the integrated intensity (area) of the elastic scattering at each $\mathbf{q}_\shortparallel$, we show in Fig. 1(d) a raw low-energy RIXS spectrum for $\mathbf{q}_\shortparallel= (0.08, 0)$ of $x=0.18$ sample, and use a Pearson VII function to fit the elastic peak (green curve) and a quadratic polynomial to account for the fluorescence background (blue curve). Figure 1(e) summarizes the on-resonance $q_\shortparallel$ scans extracted from the fitting of the elastic scattering for $x=0, 0.08, 0.11, 0.18$ (blue circles), $0.19$, and $0.21$. These on-resonance $q_\shortparallel$ scans on freshly-cleaved crystals measured at $T=15$ K reveal $q_\shortparallel$-independent flat elastic scattering, signaling the absence of 2D {\it bulk} CDW along the $H$ axis. 


However, a pronounced broad peak is unambiguously revealed in the $q_\shortparallel$ scan measured on an $x=0.18$ sample (denoted by SA18) (cyan circles in Fig. 1(e)) that was exposed to the ultra-high vacuum ($\sim10^{-10}-\sim10^{-9}$ mbar) at $T=15-20$ K for $\sim20$ hours (aging effect).
A similar peak is also observed in an aged $x=0.15 ~(T_s\approx50$ K) sample (denoted by SA15) but with an intensity $\sim1/10$ of that for the SA18 sample (Fig. 1(e)).
Figure 1(f) shows the momentum-dependent elastic scattering colormap for the SA18 sample. Two superstructure peaks are observed at symmetric positions for both positive and negative momentum transfer ($|q_\shortparallel|\approx0.10$) along $H$, revealing the unambiguous existence of a 2D periodic modulating pattern with a period of $\lambda\sim5$ nm and a correlation length of $\xi\sim4$ nm. The superstructure peak at $q_\shortparallel\approx0.10$ is strongly suppressed in the measurements with $\pi$ polarized incident photons (Fig. 1(g)). This polarization dependence (signal being enhanced with $\sigma$ polarization but suppressed with $\pi$ polarization) suggests a charge origin for the elastic scattering at $q_\shortparallel\approx0.10$. Incident-energy-dependent RIXS measurements at $q_\shortparallel=0.08$ (Fig. 1(h), (i)) and its comparison to the X-ray absorption (XAS) show that the elastic scattering is enhanced near the Fe-L$_3$ edge, demonstrating that an incommensurate CDW arises from the itinerant electrons associated with Fe $3d$ orbitals. The weak, featureless off-resonance ($E_i=706.5$ eV) scan shown in Fig. 1(j) further corroborates this identification. Figure 2(a) summarizes the on- and off-resonance $q_\shortparallel$ scans corresponding to the colormaps in Figs. 1(f), 1(g), and 1(h), presenting in the SA18 sample a clear asymmetric and broad CDW peak. 

In the STM studies revealing the smectic or stripe patterns in FeSe thin film and FeSe$_{0.81}$S$_{0.19}$, the CDWs are intimately associated with defects or vacancies on the surface \cite{Li2017, Li2021, LFA2018, LD2021, Li2022, Walker2023}. As STM is a local probe but RIXS averages over a macroscopic length scale (in our present measurements over a beam spot of 5 $\mu$m $\times$ 55 $\mu$m) with bulk sensitivity, we interpret that a low density of defects in the freshly-cleaved samples does not give rise to a CDW peak in the spatially averaged $q_\shortparallel$ scans. Thus, the emergence of the CDW peak is attributed to the aging effects that could generate more defects (or vacancies, surface reconstruction, etc.) near the surface, and the CDW associated with these defects/vacancies penetrates into the sample by a sizable depth.  The wave vector of this short-range CDW in SA18, $Q_{\rm CDW}=0.1\cdot2\pi/a\approx0.12$ {\AA}$^{-1}$, is exactly the same as that observed in the $x=0.19$ sample by STM \cite{Walker2023}, and therefore confirms the existence of a CDW in tetragonal {\FSS} by a bulk sensitive probe. 
The weaker CDW in SA15 (the nematic ordering regime) observed here by RIXS is also consistent with the absence of the stripe CDW in bulk FeSe reported in ref. \cite{Walker2023}.

To illustrate the structure factor $S(q_\shortparallel)$ of the CDW, we performed on the SA18 sample a $q_\shortparallel$ scan along $[H, H]$ direction (Fig. 2(b)) and an azimuthal dependence scan at $|q_\shortparallel|=0.08$ (Fig. 2(c)) \cite{SI}. Surprisingly, the CDW intensity at $|q_\shortparallel|=0.08$ is almost azimuthal independent, and the $q_\shortparallel$ scan along $[H, H]$ reveals a similar peak position and intensity of the CDW, indicating an isotropic charge-density correlation function in momentum space. As the CDW is observed in the NQC regime, it is natural to speculate that its emergence could be connected to strong nematic fluctuations in this region. Thus, we explore the structure factor of the CDW under uniaxial strain in order to directly infer the coupling to the electronic nematicity \cite{Chu2010, Fernandes2014}.

The mechanical uniaxial strain device is designed based on differential thermal expansion coefficients of aluminum ($\alpha\approx-24\times10^{-6}$/K), titanium ($\alpha\approx-9\times10^{-6}$/K), and invar alloy ($\alpha\approx-2\times10^{-6}$/K) to apply a temperature-dependent uniaxial strain onto a titanium platform, which can be transferred to the thin {\FSS} crystal glued on the center of the platform (Fig. 2(d)) \cite{Sunko2019}. It has been demonstrated that this strain device can apply an anisotropic (uniaxial) strain of $\varepsilon=\varepsilon_{xx}-\varepsilon_{yy}\sim-0.4\%$ to -0.8\% on {\FSS} single crystals with typical thickness $\sim10-20 ~\mu$m \cite{Liu2023}. As shown in Fig. 2(e), such a uniaxial strain can induce a sizeable spin-excitation anisotropy in the $x=0.18$ sample. As a comparison, the unstrained sample exhibits four-fold symmetric spin excitations \cite{Liu2023}. 
This anisotropy can serve as a criterion for judging that the crystal is uniaxially strained and the presence of nematic fluctuations. 

Figure 2(f) shows the $q_\shortparallel$ scans of the CDW in the uniaxial-strained SA18 sample. We found that the CDW intensity is overall enhanced under a uniaxial strain of $\varepsilon\approx-0.4$\%. More interestingly, the isotropic CDW with $Q_{\rm CDW}\approx0.10$ in the unstrained region (Fig. 2(b)), becomes highly anisotropic in uniaxial-strained SA18 (Fig. 2(f)). The $Q_{\rm CDW}$ for $H$, $K$, and $[H, H]$ directions are estimated to be $Q_h\approx0.02\pm0.04$, $Q_k\approx0.14\pm0.02$, and $Q_{hh}\approx0.06\pm0.02$. Therefore, the structure factor $S(q_\shortparallel)$ in the strained SA18 sample is elongated along the $K$ axis ($C_2$ symmetric) with the intensity distributed largely along $K$ and $[H, H]$ directions (Fig. 2(g)). Note that the isotropic $S(q_\shortparallel)$ for the unstrained SA18 sample is much akin to the isotropic CDW in electron-underdoped Nd$_{2-x}$Ce$_x$CuO$_4$ \cite{Comin2019}, which was accounted for by the topology of the Fermi surface in the underdoped regime. 


In the STM study of FeSe$_{0.81}$S$_{0.19}$ \cite{Walker2023}, the stripe patterns are locally $C_2$ symmetric with the intensity along $q_a$ or $q_b$ and the average of a large area giving $C_4$ symmetric patterns. The $S(q_\shortparallel)$ of the strained SA18 sample are qualitatively consistent with the $C_2$ symmetric stripe pattern. In the unstrained SA18, the $Q_{\rm CDW}$ is the same as that in FeSe$_{0.81}$S$_{0.19}$ \cite{Walker2023}, though it is still unclear how to reconcile the $C_4$ symmetry with the isotropic $S(q_\shortparallel)$.
Nonetheless, our results reveal an induced CDW near the surface that penetrates into the bulk of the sample by a large depth ($\sim20-100$ nm, the attenuation length of X-rays at the Fe L$_3$ edge) \cite{GA2008}. The high tunability of the symmetry and the wave vector of CDW by uniaxial strain suggests that the CDW is strongly coupled to the electronic nematicity.

\begin{figure}[htbp!]
\includegraphics[width=8 cm]{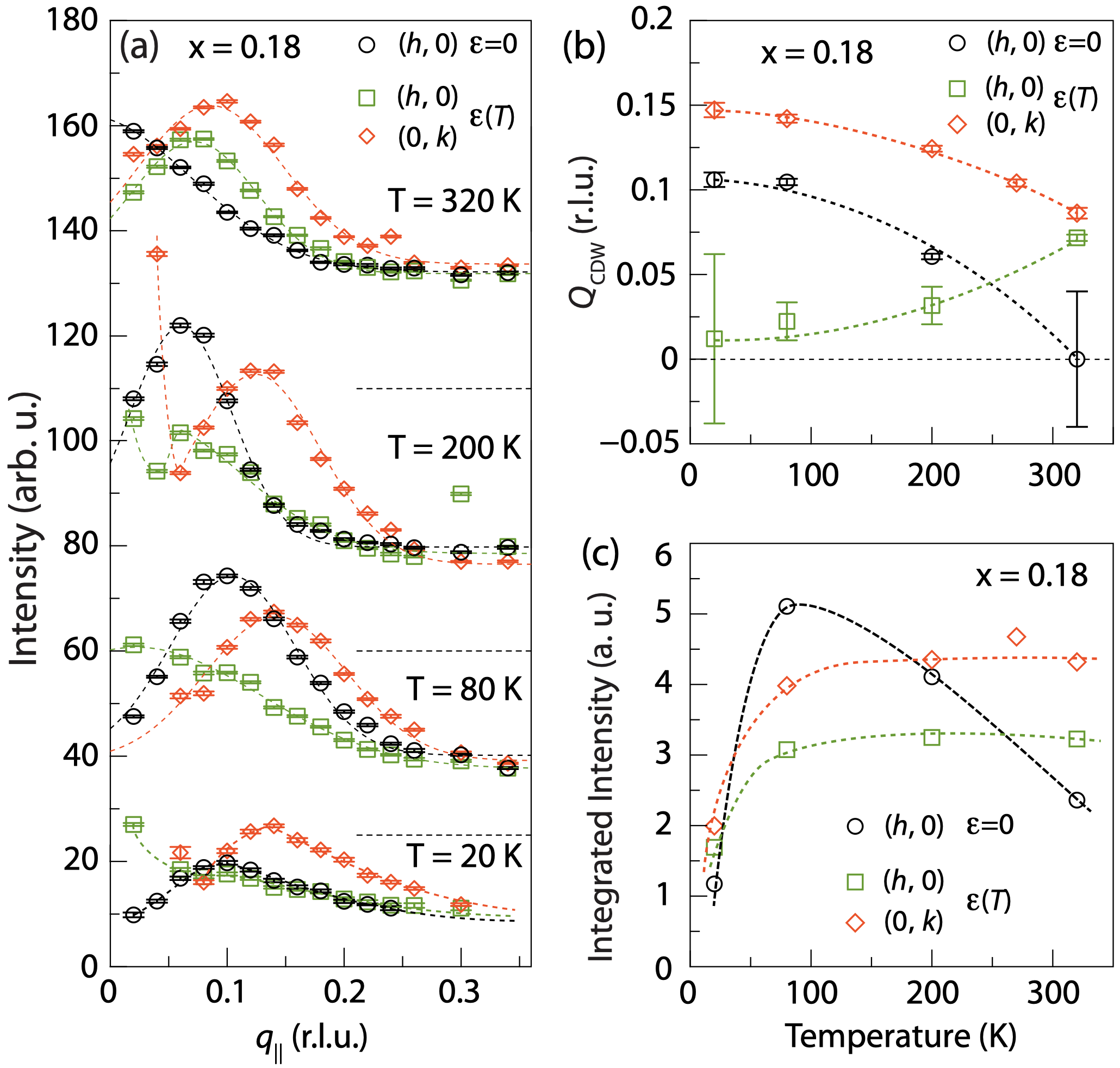}
\caption{\textbf{Temperature dependence of the CDW in FeSe$_{0.82}$S$_{0.18}$.} (a), CDW in unstrained and strained SA18 samples measured at $T = 20 K, 80 K, 200 K$, and $320$ K. (b), (c) Temperature-dependent $Q_{\rm CDW}$ and integrated intensity of the CDW peak for SA18. } 
\label{fig3}
\end{figure}

\begin{figure*}[htbp!]
\includegraphics[width=14cm]{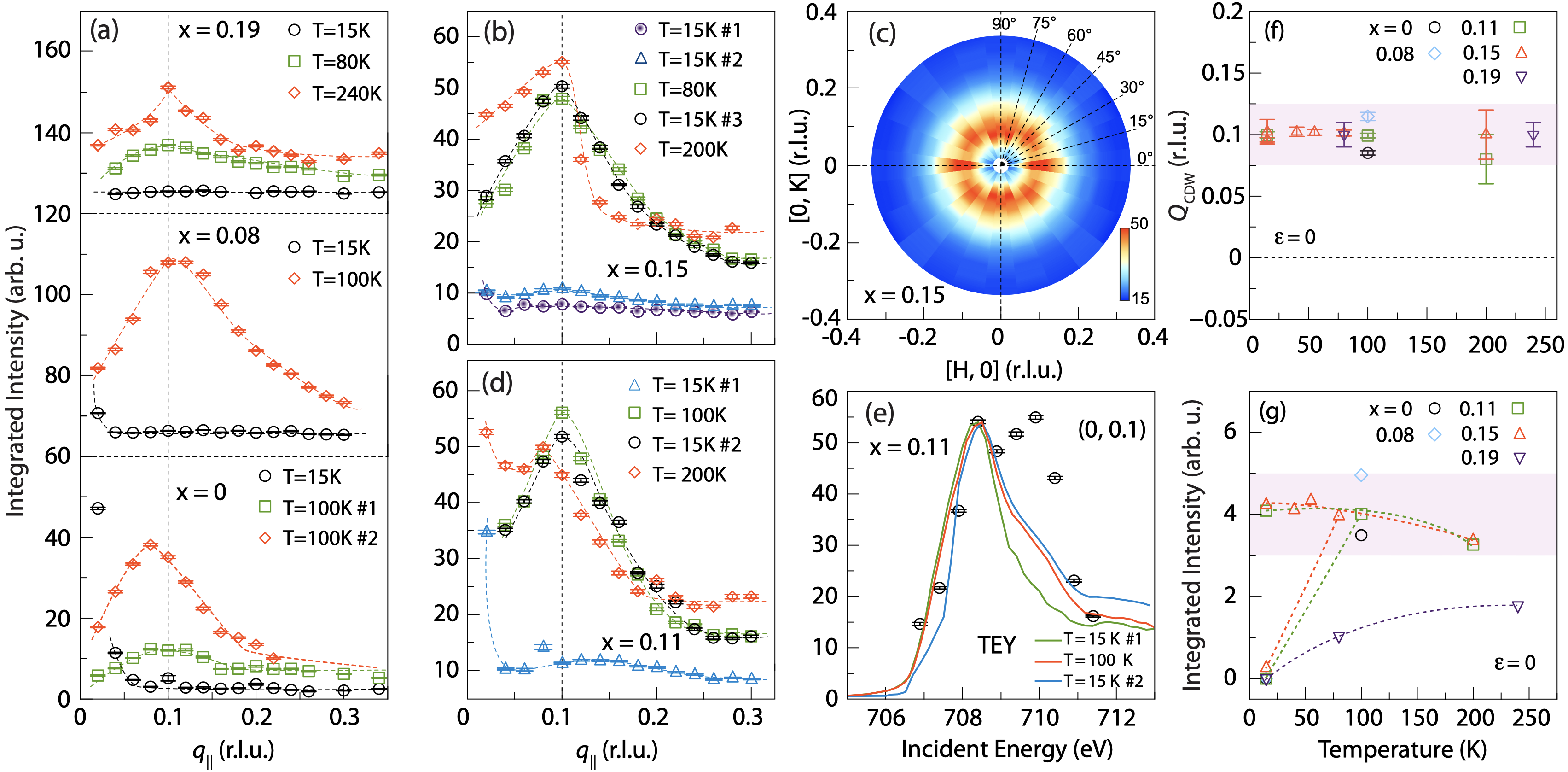}
\caption{ \textbf{Temperature and doping dependence of the CDW in {\FSS}.}
(a), (b), (d), On-resonance $q_\shortparallel$ scans for (a) $x = 0, 0.08$, and $0.19$, (b) $x=0.15$, and (d) $x=0.11$, measured at $T=15$ K, 80 K, 100 K, 200 K and 240 K. (c), Azimuthal dependence of the on-resonance $q_\shortparallel$ scans in $[H, K]$ plane for $x=0.15$. (e), The green, red and blue curves are the total-electron yield (TEY) XAS measured at $T=15$ K (\#1), $T=100$ K (measured 20 hours after $T=15$K \#1), and $T=15$K \#2 (measured 10 hours after $T=100$ K), respectively. The black circles are the integrated intensity of the incident-energy-dependent elastic scattering of $q_\shortparallel=(0.1, 0)$ measured after $T=15$K (\#2) for $x=0.11$. (f), (g) $Q_{\rm CDW}$ and integrated intensity for several dopings as a function of temperature.}
\label{fig4}
\end{figure*}

To further understand the CDW, we explore its temperature dependence in Figs. \ref{fig3} and \ref{fig4}. Figure 3(a) summarizes the temperature dependence of the CDW in unstrained and uniaxial-strained SA18. On warming to $T=80$ K, the CDW intensity in both unstrained and strained SA18 is greatly enhanced (Fig. 3(c)), while the CDW vectors remain unchanged (Fig. 3(b)) compared with those at $T=20$ K. With further increase of the temperature to $T=200$ K and $320$ K, the intensity of the CDW for unstrained $H$ direction decreases (Fig. 3(c)), and the wave vector decreases from $Q_{\rm CDW}\approx0.10$ to $Q_{\rm CDW}\approx0$ (a homogeneous charge state) (Fig. 3(b)). In comparison, the CDW for strained $H$ and $K$ directions retain their intensities at high temperatures, and their wave vectors $Q_h$ and $Q_k$ tend to merge to $Q_{\rm CDW}\approx0.10$ at $T=320$ K (Fig. 3(b), (c)). Since the uniaxial strain is gradually relaxed upon warming, the concomitant temperature- and strain-dependence of the $Q_{\rm CDW}$ in strained SA18 further demonstrate that the CDW can be continuously tuned by external uniaxial strain. The persistence of the short-range CDW at high temperatures indicates that it is not driven by a symmetry-breaking phase transition, but could be associated with certain low-energy fluctuations or short-range electron correlations.

Besides, the weak CDW in the SA15 sample depicted in Fig. 4(b) is also greatly enhanced at $T=80$ K. The CDW intensity increases gradually with time and finally reaches a saturated intensity after several hours (Supplementary Fig. S5). Once the CDW is stabilized at $T=80$ K, it becomes robust, does not change on cooling to $T=15$ K ($T=15K$ \#3 in Fig. 4(b)), and changes less on warming to $T=200$ K (Fig. 4(b)). This enhanced CDW also shows an isotropic structure factor $S(q_\shortparallel)$, as can be seen from the azimuthal-angle dependence $q_\shortparallel$ scans measured at $T=80$ K (Fig. 4(c)). Such a warming process (to $T=80-100$ K) can induce and stabilize a CDW with the same $Q_{\rm CDW}\approx0.10$ for all the dopings studied ($x=0-0.19$) (Fig. 4(a), (b), (d)). The induced CDW is also robust against the change in temperature (Fig. 4). The intensity of the CDW for $x=0.11$, and $0.15$ decreases slightly at high temperatures up to $200$ K, while that for $x=0.19$ is much weaker than the ones for lower dopings (Fig. 4(a), (g)).  

The time dependence of the CDW intensity observed in several dopings implies that certain permanent changes occur at elevated temperature with the sample, which enhances and stabilize the CDW. More importantly, as the wave vectors for all the induced (enhanced) CDWs are very close to that in the unstrained SA15 and SA18 samples, all the CDWs observed should share the same origin. 

The total electron yield (TEY) XAS spectra measured together with the $q_\shortparallel$ scans at various temperatures provide key clues to understanding the changes on the sample surface. Figure 4(e) shows the TEY XAS spectra of $x=0.11$ collected at different temperatures and times. The XAS for $T=15$ K ($T=15$K \#1) exhibits a clean sharp Fe$^{2+}$ peak at $E_i\approx708.4$ eV, while the XAS measured at $T=100$ K (35 hours after $T=15$K \#1) exhibits an apparent enhancement of the Fe$^{3+}$ signal at $E_i\approx710$ eV \cite{FeL3}, indicating the creation of some amount of Fe$^{3+}$ on the surface (Fig. 4(e)). After that, the Fe$^{3+}$ signal of the XAS remained unchanged when the sample was cooled to $T=15$ K again ($T=15$K \#2 in Fig. 4(e)) \cite{SI}. The emergence of the Fe$^{3+}$ states could be attributed to oxidation after adsorption of oxygen on the sample surface at temperatures $T\gtrsim80$ K (Supplementary Figs. S3 and S5) 
. Previous STM measurements on FeTe exposed to much higher oxygen pressure ($\sim10^{-4}$ mbar) for 15 minutes showed that oxygen could be adsorbed onto the surface of the sample but would not be substituting Te$^{2-}$ ions \cite{Ren2021}. The oxygen adsorbed on the surface could remove one electron from Fe$^{2+}$ to form a Fe$^{3+}$. 

With such Fe$^{3+}$ states, the incident-energy-dependent RIXS at $\mathbf{q}_\shortparallel$ = (0, 0.1) reveals an enhancement of the CDW signal centering at the Fe$^{3+}$ energy (black circles in Fig. 4(e)), which is absent in Fig. 1(h) for SA18 where the CDW signal is only enhanced at Fe$^{2+}$ energy. This demonstrates that the induced Fe$^{3+}$ states are also involved in the formation of the CDW. The emergence of Fe$^{3+}$ on the surface will induce local lattice distortions as it is more attractive to Se$^{2-}$/S$^{2-}$ and its ion radius ($\sim63$ pm) is obviously smaller than that of Fe$^{2+}$ ($\sim77$ pm). 
Therefore, we argue that the CDW is induced by the local lattice distortions (strains) near the surface in the electronic nematic field. This is consistent with the theoretical scenario that the``stripe-like'' smectic CDW probed by STM is caused by point- or step-like impurities on the sample surface in the electronic nematic field \cite{Lahiri2022}. However, though induced by local strains on the surface, the strong CDW intensity indicates that the penetrating depth of the CDW into the bulk should be comparable with the attenuation length of the X-rays at the Fe $L_3$ edge, which is estimated to be $\sim20-100$ nm, much larger than the penetrating depth predicted in the scenario ($1/Q_{\rm CDW} \sim 5$ nm).

We summarize below the key discoveries, discussions, and conclusions. First, the CDW observed here is neither originating from the bulk nor forming a long-range order driven by phase transition. Soft X-ray RIXS is a bulk probe with an attenuation length of typically 20-100 nm and can probe much deeper than regular surface-sensitive experimental techniques. However, this does not necessarily mean that all ordering phenomena by RIXS are originating from the bulk of the material. Second, the local strains associated with the defects (vacancies, etc.) near the surface induce and stabilize the CDW, which could penetrate into the sample by a depth much larger than $1/Q_{\rm CDW} \sim 5$ nm.
Third, the response of the CDW to uniaxial strain in SA18 indicates the CDW is intimately correlated with electronic nematicity. Thus, the CDW is induced by defect-driven local (anisotropic) strains in the electronic nematic field. These experimental facts are qualitatively consistent with the theoretical scenarios proposed in ref. \cite{Lahiri2022} but require a more quantitative understanding concerning the isotropy and the large penetrating depth of the CDW.
Our results provide new insights into understanding the interplay between electron correlations and the electronic nematicity in FeSCs. In addition, the intensive CDW that is not originating from the bulk suggests that one needs to be careful in assigning a {\it bulk} CDW in soft-X-ray RIXS/REXS measurements.



\newpage

\textbf{Methods}

The X-ray absorption (XAS) and resonant inelastic X-ray scattering measurements were performed on the RIXS spectrometer at the ADRESS beamline, Swiss Light Source, Paul Scherrer Institut \cite{ADRESS, SAXES}. We define $q_\shortparallel>0$ in the grazing-incident configuration (Fig. 1(b)) and energy loss $E=E_i-E_f$, where $E_i$ and $E_f$ are energies of the incident and scattered photons. The momentum vector $\mathbf{q}=\mathbf{k_i}-\mathbf{k_f} = (h, k, l)$ is expressed in reciprocal lattice units (r.l.u.) ($2\pi/a_o, 2\pi/b_o, 2\pi/c$).

\textbf{Acknowledgment}

The work at Beijing Normal University is supported by the National Natural Science Foundation of China (Grant Nos. 12174029, and 11922402) (X.L.). The RIXS experiments were carried out at the ADRESS beamline of the Swiss Light Source at the Paul Scherrer Institut (PSI). The work at PSI is supported by the Swiss National Science Foundation through project no. 200021\_207904, 200021\_178867, and the Sinergia network Mott Physics Beyond the Heisenberg Model (MPBH) (project numbers CRSII2 160765/1 and CRSII2 141962). Y.W. and T.C.A. acknowledge financial support from the European Union's Horizon 2020 research and innovation programme under the Marie Sk\l{}odowska-Curie grant agreement No. 884104 (Y.W.) and No. 701647 (T.C.A.) (PSI-FELLOWIII-3i).

\textbf{Author contributions}

R.L. and W.Z contributed equally to this work. 
X.L. and T.S. conceived this project, wrote the beamtime proposals, and coordinated the experiments as well as all other project phases. R.L., X.L., W.Z., Y.W., Z.T., T.C.A., and T.S. carried out the RIXS experiments with the support of V.N.S. R.L. and X.L. analysed the data. R.L. prepared the FeSe$_{1-x}$S$_x$ single crystals with help from Z. T.. X.L. and T.S. wrote the manuscript. All authors made comments.

\end{document}